
\input harvmac.tex
\def\frac#1#2{{#1\over#2}}

\def\exp{{\rm exp}}

\def\slash#1{\mathord{\mathpalette\c@ncel{#1}}}
\overfullrule=0pt

\def\steepslash{\c@ncel}
\def\frac#1#2{{#1\over #2}}

\def\inbar{\,\vrule height1.5ex width.4pt depth0pt}
\def\IB{\relax{\rm I\kern-.18em B}}
\def\IC{\relax\hbox{$\inbar\kern-.3em{\rm C}$}}
\def\IP{\relax{\rm I\kern-.18em P}}
\def\IR{\relax{\rm I\kern-.18em R}}
\def\IZ{\relax\ifmmode\mathchoice
{\hbox{Z\kern-.4em Z}}{\hbox{Z\kern-.4em Z}}
{\lower.9pt\hbox{Z\kern-.4em Z}}
{\lower1.2pt\hbox{Z\kern-.4em Z}}\else{Z\kern-.4em Z}\fi}
\catcode`\@=12

\input epsf
\noblackbox
\def\inbar{\,\vrule height1.5ex width.4pt depth0pt}
\def\IC{\relax\hbox{$\inbar\kern-.3em{\rm C}$}}
\def\IR{\relax{\rm I\kern-.18em R}}
\font\cmss=cmss10 \font\cmsss=cmss10 at 7pt
\def\IZ{\relax\ifmmode\mathchoice
{\hbox{\cmss Z\kern-.4em Z}}{\hbox{\cmss Z\kern-.4em Z}}
{\lower.9pt\hbox{\cmsss Z\kern-.4em Z}}
{\lower1.2pt\hbox{\cmsss Z\kern-.4em Z}}\else{\cmss Z\kern-.4em Z}\fi}

\font\manual=manfnt \def\dbend{\lower3.5pt\hbox{\manual\char127}}

\lref\Drin{Drinfel'd, V.G. and Sokolov, V.V.:
Lie algebras and equations of Korteweg de Vries type.
J. Sov. Math. {\bf 30}, 1975-1980 (1985) }
\lref\McCoy{McCoy, B.M.: The connection between
statistical mechanics and Quantum
Field Theory. Preprint ITP-SB-94-07, $\#$hepth 9403084 (1994);
to appear in: Bazhanov, V.V. and Burden, C.J. (eds.)
Field Theory and Statistical Mechanics. Proceedings 7-th Physics
Summer School at the Australian National University. Canberra.
January 1994, World Scientific 1995 }
\lref\Leclair{LeClair, A.: Restricted Sine-Gordon theory
and the minimal conformal series. Phys. Lett. {\bf B230}, 103-107 (1989)}
\lref\Klassen{Klassen, T.R. and Melzer, E.:
Spectral flow between conformal field theories in\ $1+1$\
dimensions. Nucl. Phys. {\bf B370}, 511-570 (1992)}
\lref\Eguchi{Eguchi, T. and Yang, S.K.:
Deformation of conformal field theories and soliton
equations. Phys. Lett. {\bf B224}, 373-378  (1989)}
\lref\Kac{Kac, V.G.:
Contravariant form for infinite-dimensional
Lie algebras and superalgebras.
Lect. Notes in Phys. {\bf 94},\ 441-445.
Berlin, Heidelberg, New York: Springer  1979}
\lref\BPZ{Itzykson, E., Saleur, H. and Zuber, J.B. (eds.)
Conformal Invariance and Applications to Statistical Mechanics,
World Scientific 1988}
\lref\DotsFat{ Dotsenko, Vl.S. and Fateev, V.A.: Conformal algebra and
multipoint correlation functions in 2d statistical models. Nucl. Phys.
B{\bf 240}\ [FS{\bf 12}], 312-348 (1984)\semi
Dotsenko, Vl.S. and  Fateev, V.A.: Four-point
correlation functions and the operator algebra in 2d conformal invariant
theories with central charge $c\le1$.
Nucl. Phys. B{\bf 251}\ [FS{\bf 13}] 691-734 (1985) }
\lref\Abf{Andrews, G., Baxter, R. and Forrester, J.:
Eight-vertex SOS model
and generalized Rogers-Ramanujan identities. J. Stat. Phys. {\bf 35},
193-266 (1984)}
\lref\Zkvadrat{Zamolodchikov, A.B. and Zamolodchikov, Al.B.:
Factorized S-matrices in two dimensions as the exact
solutions of certain relativistic quantum field theory models.
Ann. Phys. (N.Y.) {\bf 120}, 253-291 (1979) }
\lref\Fei{Feigin, B.L. and  Fuchs, D.B.  Representations
of the Virasoro algebra. In:
Faddeev, L.D., Mal'cev, A.A. (eds.) Topology.
Proceedings, Leningrad 1982.
Lect. Notes in Math.
{\bf 1060}.
Berlin, Heidelberg, New York Springer 1984}
\lref\Rks{Kulish, P.P., Reshetikhin, N.Yu. and  Sklyanin, E.K.:
Yang-Baxter equation and representation theory. Lett. Math.
Phys. {\bf 5}, 393-403 (1981) }
\lref\Dr{Drinfel'd, V.G.: Quantum Groups. In: Proceedings of
the International Congress of Mathematics. Berkeley 1986,
{\bf 1}, pp. 798-820. California: Acad. Press 1987}
\lref\Jim{Jimbo, M.: A q-difference analogue of
\ $U{\cal G}$\ and Yang-Baxter
equation. Lett. Math. Phys. {\bf 10}, 63-68 (1985)}
\lref\Fa{ Faddeev, L.D. and  Takhtajan, L.A.: Hamiltonian Method in the
Theory of Solitons. New York: Springer 1987}
\lref\Zams{ Zamolodchikov, Al.B.:
On the TBA Equations for Reflectionless ADE
Scattering Theories.
Phys. Lett. {\bf B253}, 391 (1991)}
\lref\Zand{Zamolodchikov, Al.B.:
{} From Tricritical Ising to Critical Ising by
Thermodynamic Bethe Ansatz. Nucl. Phys. {\bf B358}, 524-546 (1991)}
\lref\Gard{
Lax, P.D.: Integrals of nonlinear equations of
evolution and solitary waves. Comm. Pure Appl. Math.
{\bf 21}, 467-490 (1968)}
\lref\Miura{Miura, R.M.: Korteweg-de Vries equation and generalizations.
I. A remarkable explicit nonlinear transformation.  Phys. Rev. Lett.
{\bf 19}, 1202-1204 (1968)}
\lref\GGKM{Gardner, C.S., Greene, J.M, Kruskal, M.D. and
Miura, R.H.: Method for solving the Korteweg-de Vries equation.
Phys. Rev. Lett. {\bf 19}, 1095-1097 (1967)}
\lref\FaL{Fateev, V.A. and  Lukyanov, S.L.:
Poisson-Lie group and classical W-algebras.
Int. J. Mod. Phys. {\bf A7}, 853-876 (1992)\semi
Fateev, V.A. and  Lukyanov, S.L.: Vertex operators and
representations of quantum universal enveloping algebras.
Int. J. Mod. Phys. {\bf A7}, 1325-1359 (1992)}
\lref\Fsa{Fendley, P. and Saleur, H.:
Massless integrable quantum field theories and
massless scattering in 1+1 dimensions. Preprint USC-93-022, $\#$hepth
9310058 (1993) }
\lref\FSLS{Fendley, P., Lesage, F.  and Saleur, H.:
Solving 1d plasmas and 2d boundary problems using Jack
polynomial and functional relations. Preprint
USC-94-16, SPhT-94/107,
$\#$hepth 9409176 (1994)}
\lref\Zar{Zamolodchikov, Al.B.: Thermodynamic Bethe ansatz in
relativistic models: Scaling 3-state Potts and Lee-Yang models.
Nucl. Phys. {\bf B342}, 695-720 (1990)}
\lref\Yur{Yurov, V.P. and Zamolodchikov, Al.B.:
Truncated conformal space approach to scaling Lee-Yang model.
Int. J. Mod. Phys. {\bf A5}, 3221-3245 (1990)}
\lref\Zatr{Zamolodchikov,  Al.B.: Thermodynamic Bethe ansatz for RSOS
scattering theories. Nucl. Phys. {\bf B358}, 497-523 (1991)}
\lref\FRK{Freund, P.G.O., Klassen, T.R. and Melzer, E.:
S-matrices for perturbations of certain conformal
field theories.
Phys. Lett. {\bf B229}, 243-247 (1989)}
\lref\Sm{Smirnov, F.A.: Reductions of Quantum Sine-Gordon Model
as Perturbations of Minimal Models of Conformal Field Theory.
Nucl. Phys. {\bf B337}, 156-180 (1990)}
\lref\zkvadratdva{ Zamolodchikov, A.B. and  Zamolodchikov Al.B.:
Massless factorized scattering and sigma models with topological terms.
Nucl. Phys. {\bf B379},  602-623 (1992) }
\lref\BazR{Bazhanov, V.V. and Reshetikhin, N.Yu.:
Critical RSOS models and conformal theory.
Int. J. Mod. Phys. {\bf A4}, 115-142 (1989)}
\lref\KluP{Kl$\ddot{\rm u}$mper, A. and Pearce, P.A.:
Conformal weights of RSOS lattice models and their
fusion hierarchies. J. Phys., {\bf A183}, 304-350  (1992) }
\lref\Huse{Huse, D.A.: On the exact multicritical points in
restricted SOS models. Phys. Rev. {\bf B30}, 3908 (1984)}
\lref\YY{Yang, C.N. and Yang, C.P.: Thermodynamics of one-dimensional
system of bosons with repulsive delta-function potential.
J. Math. Phys. {\bf 10}, 1115-1123 (1969)}
\lref\Card{Cardy, J.: Boundary conditions, fusion rules and
the Verlinde formula. Nucl. Phys. {\bf B324}, 581-596 (189)}
\lref\Cards{Cardy, J.  Conformal invariance and surface critical behavior.
In: Itzykson., E., Saleur, H. and Zuber, J.B. (eds.)
Conformal Invariance and Applications to Statistical Mechanics,
World Scientific 1988}
\lref\ABZ{Zamolodchikov, A.B.:
Integrable field theory from conformal field theory.
Adv. Stud. in Pure Math.
{\bf 19}, 641-674 (1989)}
\lref\GZ{Ghosal, S. and Zamolodchikov, A.B.:
Boundary S-matrix and boundary state in two-dimensional integrable
Quantum Field theory.
Int. J.  Mod. Phys. {\bf A9}, 3841-3885 (1994)}
\lref\FadS{Faddeev, L.D., Sklyanin, E.K. and Takhtajan, L.A.:
Quantum inverse scattering method I.
Theor. Math. Phys. {\bf 40}, 194-219 (1979) (in Russian)}
\lref\Kup{Kupershmidt, B.A. and Mathieu, P.:
Quantum KdV like equations and perturbed
Conformal Field theories. Phys. Lett. {\bf B227}, 245-250  (1989)}
\lref\Ver{Verlinde., E.:
Fusion rules and modular transformations in 2d Conformal Field
Theory. Nucl. Phys.  {\bf B300}, 360-376 (1988)}
\lref\dVega{de Vega, H.J. and Destri., C.:
Unified approach to thermodynamic Bethe Ansatz and finite size corrections
for lattice models and field theories. Preprint
IFUM 477/FT, LPTHE 94-28, $\#$hepth 9407117 (1994)}
\lref\jap{Sasaki, R. and Yamanaka, I.:
Virasoro algebra, vertex operators, quantum Sine-Gordon
and solvable Quantum Field theories. Adv. Stud. in Pure Math.
{\bf 16}, 271-296 (1988) }
\lref\BASZ{Bazhanov, V.V., Lukyanov, S.L. and Zamolodchikov, A.B.:
In preparation. }
\lref\Kor{Bogoliubov, N.M., Izergin, A.G. and  Korepin, V.E.:
Correlation functions in integrable systems and the
Quantum Inverse Scattering Method.
Moscow: Nauka 1992 (In Russian)}
\lref\Fendly{Fendley, P.: Exited state thermodynamics.
Nucl. Phys. {\bf B374}, 667-691 (1992)}
\lref\MKlas{Kedem, R., Klassen, T.R., McCoy B.M. and Melzer E.:
Fermionic sum representations for conformal field theory
characters. Phys. Lett. {\bf B307}, 68-76 (1993)}
\lref\Zamt{Zamolodchikov, Al.B.: Private communication.}
\lref\Baxter{Baxter, R.J.: Exactly Solved Models
in Statistical Mechanics. London: Academic Press 1982}
\lref\KR{Kirilov, A.N. and Reshetikhin N.Yu.:
Exact solution of the integrable $ XXZ$  Heisenberg model with
arbitrary spin. J. Phys. {\bf A20}, 1565-1585 (1987)}
\lref\BP{Baxter, R.J. and Pearce, P.A.: Hard hexagons:
interfacial tension and correlation length. J. Phys.
{\bf A15}, 897-910 (1982)}

\Title{\vbox{\baselineskip12pt\hbox{CLNS  94/1316}
\hbox{RU-94-98}
\hbox{hep-th/9412229}}}
{\vbox{\centerline
{Integrable Structure of Conformal Field Theory,}
\vskip6pt\centerline{Quantum KdV
Theory}
\vskip6pt\centerline{and Thermodynamic Bethe Ansatz}}}

\centerline{Vladimir V. Bazhanov$^1$\footnote{$^*$}
{e-mail address: Vladimir.Bazhanov@maths.anu.edu.au}\footnote{$^{\sharp}$}
{On leave of absence from the Institute for High Energy Physics,}
\footnote{}
{Protvino, Moscow Region, 142284, Russia},
Sergei L. Lukyanov$^2$ \footnote{$^{**}$}
{e-mail address: sergei@hepth.cornell.edu}}
\centerline{
and Alexander B. Zamolodchikov$^3$ \footnote{$^\dagger$}
{e-mail address: sashaz@physics.rutgers.edu} }

\centerline{}
\centerline{}

\centerline{$^1 $ Department of
Theoretical Physics and Center of Mathematics}
\centerline{and its Applications, IAS, Australian National University, }
\centerline{Canberra, ACT 0200, Australia}
\centerline{$^2$Newman Laboratory, Cornell University}
\centerline{ Ithaca, NY 14853-5001, USA}
\centerline{$^3$Department of Physics and Astronomy,}
\centerline{Rutgers University, Piscataway, NJ 08855-049, USA}
\centerline{and}
\centerline{L.D. Landau Institute for Theoretical Physics,}
\centerline{Chernogolovka, 142432, Russia}
\centerline{}
\centerline{}

\Date{December, 94}

\eject
\vbox{
\centerline{{\bf Abstract}}
\centerline{}
We construct the quantum versions of the monodromy matrices of KdV
theory. The traces of these quantum monodromy matrices,
which will be called as ``${\bf T}$-operators'',
act in  highest weight Virasoro modules.
The ${\bf T}$-operators depend on the spectral parameter $\lambda$ and
their expansion around $\lambda = \infty$ generates an infinite set of
commuting Hamiltonians of the quantum KdV system. The ${\bf T}$-operators
can be viewed as the continuous field theory versions of the commuting
transfer-matrices of integrable lattice theory. In particular, we show
that for the values $c=1-3{{(2n+1)^2}\over {2n+3}}
\ ,\ \ n=1,2,3...\  $ of the Virasoro central charge the
eigenvalues of the ${\bf T}$-operators satisfy a closed system of
functional equations sufficient for determining the spectrum. For the
ground-state eigenvalue these functional equations are equivalent to those
of massless Thermodynamic Bethe Ansatz for the minimal
conformal field theory ${\cal M}_{2,2n+3}$; in general they provide a
way to generalize the technique of Thermodynamic Bethe Ansatz to the
excited states. We discuss a generalization of our approach to the cases
of massive field theories obtained  by  perturbing  these Conformal
Field Theories with the operator $\Phi_{1,3}$. The relation of these ${\bf
T}$-operators to the boundary states is also briefly described.}

\eject

The studies of the last decade revealed a deep relation between the
structures of Conformal Field Theory (CFT)\ \BPZ ,\
Integrable Field Theory\ \FadS ,\ \Kor\  and Solvable Lattice Models
\ \Baxter . The
conformal symmetry of CFT is generated by its energy-momentum tensor
$T(u)$, whose mode expansion

\eqn\ten{T(u)=-{c\over 24}+\sum_{-\infty}^{+\infty}L_{-n}\ e^{inu} }
is expressed in terms of the operators $L_n$ satisfying the
commutation relations

\eqn\iaj{[L_n, L_m]=(n-m)\ L_{n+m}+{c\over 12}(n^3-n)\ \delta_{n+m,0}}
of Virasoro algebra $Vir$\ \BPZ .
Here $c$ is the central charge which is the
most important characteristic of CFT;
in writing \ \ten\  we have chosen the
periodic boundary conditions $T(u+2\pi)=T(u)$. The ``chiral'' space of
states ${\cal H}_{chiral}$ is built up from a collection of
suitably chosen irreducible highest weight modules \footnote{$^1$}{Of
course the physical space of states ${\cal H}_{phys}$ is embedded (by
diagonal or some other suitable embedding) into a tensor product ${\cal
H}_{chiral}\otimes {\bar{\cal H}}_{chiral}$, see e.g.\  \BPZ .},

\eqn\jdgt{{\cal H}_{chiral}=\oplus_{a}\  {\cal V}_{a}\ ,}
where\ ${\cal V}_a\equiv {\cal V}_{\Delta_a}$\ , the parameters
$\Delta$ are the highest weights, and associated highest weight vectors
$\mid \Delta \rangle \in {\cal V}_a$ satisfy the equations

\eqn\mbnb{
L_n \mid \Delta \rangle=0\quad {\rm for}
\quad n>0\ ; \qquad L_0\mid \Delta \rangle=
\Delta\mid \Delta\rangle\ . }

Along with this ``conventional'' characterization of CFT a quite
different description of CFT in terms of ``massless S-matrix'' has been
recently proposed\ \Zand ,\ \zkvadratdva  ,\ \Fsa .
In this approach the chiral states of CFT are
described as the scattering states of a collection of ``massless
particles'' and the CFT is characterized \footnote{$^2$}{Unlike the case
of massive field theory the ``factorizable S-matrix'' description of CFT
in general is not unique; there can exist several different choices
\ \McCoy\  of
the ``massless particle'' basis in ${\cal H}_{chiral}$, with different
particle contents and scattering amplitudes, which correspond to massless
limits of different integrable perturbations of this CFT .} by their
factorizable S-matrix, rather then in terms of Virasoro algebra and
its highest weight representations. So far the relation between these
two description is understood to a very limited extent. The ``massless
S-matrix'' allows one to compute, through the Thermodynamic Bethe
Ansatz technique\ \YY ,\
\Zar , the asymptotic density of states and thus gives
the central charge $c$ (and sometimes few of the highest weights $\Delta$
\ \Fendly).
In general case even the correspondence between the ``massless
particle states'' and vectors in \ \jdgt\  is not known. At the same time
gaining more understanding about this relation is very important as the
``massless S-matrix'' description seems to capture much of the
``integrable structure'' of CFT, i.e. it is
close to its
description  in terms of  action-angle variables.

{}From the algebraic point of view these ``massless particle states'' are
nothing else then the eigenstates of an infinite set of commuting
Integrals of Motion\ (IM). It has been known for some while that if we
consider the algebra $UVir$ generated by the energy-momentum tensor \ \ten\
along with various ``composite fields'' built as a powers of $T(u)$ and
its derivatives, this algebra contains an infinite-dimensional abilean
subalgebra \ \jap ,\ \Eguchi\
spanned by the
``local IM'' $I_{2k-1}\in UVir ,\  k=1,2,...$ which have the form

\eqn\inas{I_{2k-1}=\int_{0}^{2\pi}{du\over{2\pi}}\ T_{2k}(u)\ ,}
where the densities $T_{2k}(u)$ are appropriately regularized
polynomials in $T(u)$ and its derivatives. The first few densities
$T_{2k}(u)$ can be written as

\eqn\qda{T_2 (u) = T(u)\ , \qquad  T_4 (u) = : T^2 (u):\ , \qquad
T_6 (u) = :T^3 (u): + {{c+2}\over 12} :(T'(u))^2 :\ ,  \  ... }
Here the prime stands for the derivative and $:\quad :$ denotes
appropriately regularized products of the fields, for example

\eqn\kdi{:T^2 (u): =\oint_{\cal C}
{dw\over{2\pi i}}\ {1\over{w-u}}{\cal T}
\big(T(w)T(u)\big)\ ,}
where the symbol $\cal T$ denotes the ``chronological ordering'', i.e.

\eqn\lki{
{\cal T}\big(A(w)B(u)\big)=\cases{
A(w)B(u)\ ,\ \    & if\ \ \  $  \Im m\  u > \Im m\  w$\ ;\cr
B(u)A(w)
\  ,\ \  & if\ \ \  $  \Im m\  w > \Im m\  u$\ .\cr}}
In writing\ \qda\ we disregarded    all the terms which
are total derivatives and do not contribute to \ \inas .
Although a general expression for all densities $T_{2k}(u)$ is not
known  they are uniquely determined (up to a normalization which we
will fix later) by the requirement of the commutativity

\eqn\kdjfh{[I_{2k-1},I_{2l-1}]=0}
and the ``spin assignment'' \footnote{$^3$}{Another way to formulate this
requirement is to assign the grade 2 to the field $T(u)$ and the grade 1
to the derivative; then the density $T_{2k} (u)$ is a homogeneous
polynomial in $T(u)$ and its derivatives of the total grade $2k$.}

\eqn\kfu{\oint_{\cal C}{dw\over{2\pi i}}\
(w-u){\cal T}\big(T(w)T_{2k}(u)\big)
=2k\ T_{2k}(u)\ .}
More representatives (beyond \ \qda\ ) of this infinite set of
densities $T_{2k}(u)$
can be found in \ \jap . The integrals \ \inas\  define
operators $I_{2k-1}: {\cal V}_{\Delta} \to {\cal V}_{\Delta}$ which can be
expressed in terms of the
Virasoro generators $L_n$, for example

\eqn\jdufh{\eqalign{&I_1 = L_0 - {c\over 24}\ , \cr
I_3=2\ \sum_{n=1}^{+\infty}L_{-n}L_{n}&+L_0^2
-{c+2\over 12}\ L_0+{c\ (5 c+22)\over 2880}\ ,
\cr I_5 = \sum_{n_1+n_2+n_3=0}:L_{n_1}L_{n_2}L_{n_3}:&+
\sum_{n=1}^{+\infty}{\Bigl({c+11\over 6}\  n^2
-1-{c\over 4}}\Bigl) \
L_{-n}L_{n}+\cr
{3\over 2}\ \sum_{r=1}^{+\infty}\  L_{1-2 r}L_{2 r-1} -{c+4\over 8}\
 L_0^2&+{(c+2)\ (3c+20)\over 576}\ L_0-
{c\  (3 c+14)\  (7 c+68)\over 290304} \ ,\cr
&\ \ \ \ \ \ \ \ \ ...\ .}}
The ``normal ordering''
$:\quad :$ in these formulas
means that the operators $L_n$ with the bigger $n$ are
placed to the right.
Note that although these operators are not polynomial in $L_n$ their
actions in $ {\cal V}_{\Delta}$ are well defined.

In this work we study the problem of simultaneous diagonalization of the
operators $I_{2k-1}$ in ${\cal V}_{\Delta}$ by using the
approach which can be regarded as a version of the
Quantum Inverse Scattering
Method\ \FadS ,\ \Kor . It was remarked many times
\ \jap ,\ \Eguchi , \Kup\  that this
problem can be thought of as the quantum version of the  KdV problem as it
reduces to the classical KdV problem (with periodic boundary conditions)
in its ``classical limit'' $c\to -\infty$. Indeed, in this limit the
substitution

$$T(u)\to -\  {c\over 6}\ U(u)\ ,\ \ \ \ \ [\ ,\ ]\to\  { 6 \pi\over i c}
 \  \{\ ,\ \}$$
($U(u+2\pi)=U(u)$) reduce
the algebra\ \ten , \ \iaj\  to the Poisson bracket
algebra

\eqn\pois{\lbrace U(u),U(v)\rbrace
= 2\ \big(U(u)+U(v)\big)\  \delta'(u-v)+\delta'''(u-v)\ ,}
which is well known to describe the
second Hamiltonian structure of the KdV
equation provided we take one of the infinite set of classical IM
$I_{2k-1}^{(cl)}$

\eqn\kis{\eqalign{&I_{1}^{(cl)}=\int_{0}^{2\pi}{du\over{2\pi}}\  U(u)\ ,\cr
&I_{3}^{(cl)}=\int_{0}^{2\pi}{du\over{2\pi}}\  U^{2}(u)\  ,\cr
&I_{5}^{(cl)}=
\int_{0}^{2\pi}{du\over{2\pi}}\  \big[U^{3}(u)-{(U'(u))^2\over 2}
\big]\ \cr
&\ \ \ \ \ \ \ \ \ \ \ \ \ \ \ \ \ \ \   ...\ ,}}
as the Hamiltonian. The classical IM \  \kis\ , which form a commutative
Poisson bracket algebra $\lbrace
I_{2k-1}^{(cl)},I_{2l-1}^{(cl)}\rbrace=0$, evidently are the
classical versions of the operators \ \inas ,\ \jdufh .

It is also well known\ \Gard \ that the KdV equations
\eqn\pot{\partial_{t_{2k-1}}
U(t_1,t_3,...)=\{I_{2k-1}, U(t_1,t_3,..)\}\ ,\ \
t_1
\equiv u}
describe isospectral
deformations of the second order differential operator

\eqn\lop{L=\partial_u^2+U(u)-\lambda^2\ .}
In particular, if we define the $2\times 2$ Monodromy Matrix
${\bf M}(\lambda)$,
which belongs to the group\ $SL(2)$, as

\eqn\matr{\big(\psi_1 (u+2\pi), \psi_2 (u+2\pi)\big ) =
\big(\psi_1 (u), \psi_2 (u)\big )\  {\bf
M}(\lambda )\ ,}
where $\psi_1 (u),\psi_2 (u)$ are two linearly independent solutions to
the equation $L\psi=0$, then the eigenvalues of ${\bf M}(\lambda)$ are
involutive (with respect to the Poisson structure \ \pois )  Integrals of
Motion of the KdV flows, and the trace

\eqn\tre{{\bf T}(\lambda)=tr\  {\bf M}(\lambda)}
can be thought of as the
generating function for the local IM \ \kis\  as it
expands in the asymptotic series

\eqn\qexp{{1\over 2\pi}\ \log\big[ {\bf  T}(\lambda)\big] \simeq \lambda-
\sum_{n=1}^{\infty}\ c_n\  I_{2n-  1}^{(cl)}\
{\lambda^{1-2n}}\  ,}
here \ $c_1={1\over 2}\ ,\ \ c_n={(2n-3)!!\over 2^n n!}\ , n>1\ .$

It is more convenient for our purposes to start with the first order
differential operator

\eqn\matro{{\cal L}=\partial_u-\phi'(u)\ H -\lambda\  (E+F)\ ,}
where $\phi(u)$ are the canonical variables with the Poisson
brackets

\eqn\skob{\eqalign{\lbrace \phi(u),\phi(v)\rbrace =&
 \ {1\over 2}\ \epsilon (u-v)\ ;\cr
\epsilon(u)= n \qquad {\rm for}& \quad 2\pi n < u < 2\pi(n+1)\ ;
 \quad n \in {\bf  Z}\ ,}}
which are related to $U(u)$ by Miura transform\ \Miura

\eqn\miura{U(u)=-\phi'(u)^2 - \phi''(u)\ ,}
and $E,F,H$ are the generators of the Lie algebra $sl(2)$,

\eqn\lia{[H,E]=2E,\quad [H,F]=-2F, \quad[E,F]=2H\ .}
In general, the classical field $\phi(u)$ has to be taken quasiperiodic

\eqn\per{\phi(u+2\pi)=\phi(u)+2\pi i p}
to guarantee the periodicity of $U(u)$. In order to define the monodromy
matrix for the operator \ \matro\  one has
to pick some matrix representation for the $sl(2)$ algebra \ \lia . Let
$\pi_j[E], \pi_j[F], \pi_j[H]$ be $(2j+1)\times (2j+1)$ matrices,
$j=0,1/2,1,3/2,...$, representing\  \lia , such that
$\pi_j[H]=diag(2j,2j-1,...,-2j)$. Then, for given $j$, the solution to
the equation ${\cal L}\Psi(u)=0$ is
\eqn\sol{\Psi(u)=\pi_j\Bigl[e^{\phi(u) H}{\cal P} \exp\big(\lambda
\int_{0}^{u}dv(e^{-2\phi(v)} E+ e^{2\phi(v)} F)\big)\Bigl]\Psi_0\ ,}
where the symbol ${\cal P}$ denotes the ``path ordered'' exponential and
$\Psi_0$ is arbitrary vector in ${\bf C}^{2j+1}$. The associated
monodromy matrices have the form
\eqn\ku{{\bf\ M}_j (\lambda)=\pi_j\Bigl[e^{2\pi i pH}{\cal P}
\exp\big(\lambda \int_{0}^{2\pi}dv
(e^{-2\phi(v)} E+ e^{2\phi(v)} F )\big)\Bigl]\ .}
Let us introduce  auxiliary matrices
\eqn\lapo{{\bf L}_j(\lambda)=\pi_j[e^{-\pi i p H}]\  {\bf M}_j
(\lambda)\ .}
They satisfy the ``$r$-matrix''
Poisson bracket algebra\ \Fa

\eqn\alfrse{\lbrace {\bf L}_j (\lambda)\matrix{{}\cr\otimes\cr{} {^{,}}}
{\bf L}_{j'}(\mu)\rbrace =
[{\bf r}_{j j'}({\lambda \mu^{-1}}),{\bf L}_j (\lambda)\otimes
{\bf L}_{j'}(\mu)]\ ,}
where ${\bf r}_{j  j'}(\lambda)=\pi_j\otimes\pi_{j'}[\ {\bf  r}\ ]$
is the ``classical $r$-matrix''

\eqn\jhsg{{\bf r}(\lambda) ={\lambda+\lambda^{-1}\over\lambda-\lambda^{-1}}
{H\otimes H\over2}+{2\over \lambda-\lambda^{-1}}
\big(E\otimes F +F\otimes E\big)\ .}
It follows immediately that the quantities

\eqn\comm{{\bf T}_j (\lambda)=tr\ {\bf M}_j (\lambda)}
are in involution with respect to the Poisson bracket \ \pois ,\ \skob  :

\eqn\gt{\lbrace {\bf T}_j (\lambda), {\bf T}_{j'} (\mu)\rbrace =0\ .}
In particular, ${\bf T}_{1\over 2}(\lambda)$ coincides with \ \tre .

After this brief review of known classical results we turn to the
quantum case. Quantum version of the Miura transform \ \miura\   is the
Feigin-Fuchs ``free field representation'' of $Vir$\ \Fei ,\  \DotsFat

\eqn\qmiura{-\beta^2 \ T(u)
=:\varphi'(u)^2 : +(1-\beta^2) \varphi''(u)+{\beta^2\over
24}\ ,}
in terms of the free field operator

\eqn\fil{\varphi(u)=
i   Q + i  P u +\   \sum_{n\not=0}
{a_{-n}\over n} e^{in u}\ ,}
where the mode operators $Q, P$ and $a_n$ satisfy the Heisenberg algebra

\eqn\hei{[Q,P]={i\over 2}\ \beta^2 ;
\qquad [a_n,a_m]={n\over 2}\   \beta^2\ \ \delta_{n+m,0}}
($P$ and $Q$ commute with $a_n$)
and the parameter $\beta$ is  related to the central charge  $c$ as

\eqn\jsudg{\beta=\sqrt{{1-c \over 24}}-\sqrt{{25-c\over 24}}\ .}

Let ${\cal F}_p$ (``Fock space'') be the highest weight module over the
Heisenberg algebra \ \hei\  with the highest weight vector $\mid p
\rangle$ (``vacuum'') defined by the equations

\eqn\nvb{P\mid p \rangle =
p\mid p \rangle ;\qquad a_n\mid p \rangle=0\quad
{\rm for}\quad n>0\ .}
The Eq.\ \qmiura\  defines the action of $Vir$ in ${\cal F}_p$.
For generic $c$ and $p$ the space ${\cal F}_p$ is isomorphic to the
highest weight Virasoro module ${\cal V}_{\Delta}$ with\ \Fei

\eqn\ras{\Delta=\Bigr(\frac{p}{\beta}\Bigr)^2+{{c-1}\over 24}\ .}
The space ${\cal F}_p$ naturally splits into the sum of
finite-dimensional ``level subspaces''

\eqn\fockl{{\cal F}_p = \oplus_{l=0}^{\infty} {\cal F}_{p}^{(l)}\ ;
\qquad L_0\  {\cal F}_{p}^{(l)} = (\Delta + l)\  {\cal F}_{p}^{(l)}.}
The ``normal ordering'' suitable for this representation is implied in
\ \qmiura , $:\quad :$
means that the operators $a_n$ with the bigger $n$ are
placed to the right. Of course the IM \ \inas , \ \jdufh\
can be expressed in
terms of the Feigin-Fuchs free field $\varphi'(u)$. We adopt the following
normalization of the operators $I_{2k-1}$:
\eqn\pol{I_{2k-1}=(-1)^k\beta^{-2k}\
\int_{0}^{2\pi}{du\over 2\pi}\Bigr(:\big(\varphi'(u)\big)^{2k}: +
.. \Bigr)\ ,}
where the omitted terms contain higher derivatives of $\varphi (u)$.
These operators act invariantly in the level subspaces
${\cal F}_{p}^{(l)}$. Therefore diagonalization of $I_{2k-1}$ in a given
level subspace reduces to a finite algebraic problem which however
rapidly becomes very complex for higher levels. Here we list only the
vacuum eigenvalues for the first few $I_{2k-1}$,

\eqn\lsoi{I_{2k-1}\mid p \rangle = I_{2k-1}^{(vac)}(\Delta)
\mid p \rangle\ ,}
where
\eqn\nxbcvc{\eqalign{
&I_1^{vac}(\Delta)=\Delta-\frac{c}{24}\ ,\cr
&I_3^{vac}(\Delta)=\Delta^2- \frac{c+2}{12}\ \Delta+
\frac{c\ (5c+22)}{2880}\ ,\cr
&I_5^{vac}(\Delta)=\Delta^3-\frac{c+4}{8}\ \Delta^2+
\frac{(c+2)\ (3c+20)}{576}\
\Delta-\frac{c\  (3c+14) \ (7c+68)}{290304}\ ,\cr
&I_7^{vac}(\Delta)=\Delta^4-\frac{c+6}{6}\ \Delta^3+
\frac{15  c^2+194  c+568}{1440}\ \Delta^2-\cr
&\ \ \ \ \ \ \ \ \ \ \ \ \ \ \ \ \
\frac{(c+2)\  (c+10)\ (3 c+28)}{10368}\ \Delta
+\frac{c\  (3 c+46)\  (25 c^2+426 c+1400) }{24883200}\ .}}



After these preparations we can define a quantum version of the
Monodromy Matrices \ \ku\
and the operators \ \lapo , \comm ; we
are going to use the same symbols\ ${\bf L}_j,\
{\bf T}_j$\ for the quantum counterparts of \ \lapo , \comm .
Consider the following
operator valued matrices\ \FaL

\eqn\loi{{\bf L}_j (\lambda) = \pi_j \bigg [e^{i \pi P H}
 {\cal P}\exp\big(\lambda
\int_{0}^{2\pi}du (:e^{-2\varphi (u)}:q^{H\over 2}E + :e^{2\varphi
(u)}:q^{-{H\over 2}}F)\big) \bigg ]\ ,}
where the vertex operators

\eqn\sr{:e^{\pm 2\varphi (u)}
:\qquad : \qquad {\cal F}_p \to {\cal F}_{p\pm
\beta^2} }
are defined as

\eqn\mcjf{:e^{\pm 2 \varphi (u)}: \equiv
\exp(\pm 2 \sum_{n=1}^{\infty}{a_{-n}\over
n}e^{inu})\ \exp\big(\pm 2  i\ (Q+Pu)\big)\ \exp(\mp 2
\sum_{n=1}^{\infty}{a_{n}\over
n}e^{-inu})\ ,}
and $E, F$ and $H$ are the generating elements of the quantum universal
enveloping algebra  $U_q (sl(2)) $ \Rks

\eqn\uals{[H,E]=2E,
\qquad [H,F]=-2F, \qquad [E,F]={{q^{H}-q^{-H}}\over {q -
q^{-1}}}}
with

\eqn\qpar{q=e^{i\pi \beta^2}\ .}
The symbol $\pi_j$ in \ \loi\  stands again for the $(2j+1)$
dimensional representation of $U_q (sl(2))$ so that \ \loi\  is in fact
$(2j+1)\times (2j+1)$ matrix whose elements are the operators in

\eqn\nxbd{\hat {\cal F}_{p}
 = \oplus_{n=-\infty}^{+\infty}{\cal F}_{p+ n\beta^2}\ .}
These operators are understood as the power series in $\lambda$,

\eqn\troi{{\bf L}_j (\lambda)=
\pi_j \bigg [ e^{i\pi P H} \sum_{k=0}^{\infty}
\lambda^k \int_{2\pi\geq u_1 \geq u_2 \geq ... \geq
u_k\geq 0}K(u_1)K(u_2)...K(u_k)\ du_1 du_2 ... du_k\bigg ]\ ;}

$$K(u)= :e^{-2\varphi (u)}: q^{H\over 2}E+:e^{2\varphi (u)}: q^{-{H\over
2}}F\ . $$
The integrals in\  \troi\  are convergent for

\eqn\cent{-\infty < c < -2\ .}
In fact, the operators \ \loi\  can be defined for wider range of $c$ by
appropriate regularization of divergent
integrals in \ \troi . In this paper
we restrict our attention to the domain \ \cent .

The operator matrices \ \loi\  are
designed in such a way that they satisfy
the Quantum Yang-Baxter Equation

\eqn\yab{{\bf R}_{jj'}(\lambda\mu^{-1})\
\big({\bf L}_j(\lambda)\otimes 1\big)\
\big(1\otimes {\bf L}_{j'}(\mu)\big)=
\big(1\otimes {\bf L}_{j'}(\mu)\big) \
\big({\bf L}_{j}(\lambda)\otimes 1\big)\ {\bf R}_{jj'}
(\lambda \mu^{-1})\ ,}
where the matrix \ ${\bf R}_{jj'}(\lambda)$\ is trigonometric solution
of the Yang-Baxter
equation which acts in the space\ $\pi_j\otimes\pi_{j'}$. In particular
\eqn\bxvc{\eqalign{{\bf R}_{{1\over 2} {1\over 2}}(\lambda)
=\pmatrix{q^{-1}\lambda-q \lambda^{-1}&
{}&{}&{}\cr
{}&\lambda-\lambda^{-1}&q^{-1}-q&{}\cr
{}&q^{-1}-q&\lambda-\lambda^{-1}&{}\cr
{}&{}&{}&q^{-1}\lambda-q \lambda^{-1}}\ .}}
The validity of \ \yab\  can be checked
explicitly for the first few orders of the expansion of \ \yab\  in
$\lambda$ and $\mu$ with the
use of \ \troi . One can prove \ \yab\  to all
orders by taking the discrete approximations to the ${\cal P}$-ordered
integral in \ \loi\ \FaL .

Let us define now the operators

\eqn\mjvb{{\bf T}_j (\lambda) =
 tr_{\pi_j}\big(\ e^{i\pi PH}\  {\bf L}_j (\lambda)\ \big)\ ,}
which satisfy

\eqn\mcn{[{\bf T}_j (\lambda),{\bf T}_{j'} (\mu)]=0}
as a simple consequence of \ \yab\ . It is easy to see that the operators
\ \mjvb\  commute with the operator $P$ and hence they act invariantly in
${\cal F}_p$. Moreover, one can check by direct calculations that

\eqn\nsf{[{\bf T}_j (\lambda), I_{2k-1}]=0\ .}
It follows, in particular, that the level subspaces ${\cal F}_{p}^{(l)}$
are the eigenspaces of ${\bf T}_j (\lambda)$.

Let us concentrate first on the simplest nontrivial ${\bf T}$-operator
${\bf T}(\lambda)={\bf T}_{1\over 2}(\lambda)$ which corresponds to the
two-dimensional representation of $U_q (sl(2))$. In this case

\eqn\rew{\pi_{1\over 2}(H)=\bigg ( {{1 \ \ \ 0}\atop {0 -1}} \bigg );\qquad
\pi_{1\over 2}(E)=\bigg ( {{0 \ \ \ 1}\atop {0 \ \ \ 0}} \bigg );\qquad
\pi_{1\over 2}(F)=\bigg ( {{0 \ \ \ 0}\atop {1 \ \ \ 0}} \bigg )\ . }
Substituting \ \troi\  into \ \mjvb\  and computing the trace one finds

\eqn\bcvf{
{\bf T}(\lambda)=2\cos (2\pi P) +
 \sum_{n=1}^{\infty}\lambda^{2n}\ Q_n\ ,}
where

\eqn\plo{\eqalign{
Q_n =q^n &\int_{2\pi\geq u_1 \geq u_2 \geq ... \geq u_{2n} \geq 0}
\big
(e^{2 i\pi P}:e^{-2\varphi(u_1)}::e^{2\varphi(u_2)}:
:e^{-2\varphi(u_3)}: ...
:e^{2\varphi(u_{2n})}:+\cr
&e^{-2 i\pi P}:e^{2\varphi(u_1)}: :e^{-2\varphi(u_2)}:
:e^{2\varphi(u_3)}: ... :e^{-2\varphi(u_{2n})}: \big)\ du_1...du_{2n}\ .}}
So, the operator ${\bf T}(\lambda)$ can be considered as the generating
function for the ``nonlocal IM'' $Q_n$ which commute among themselves

\eqn\ki{[Q_n , Q_m]=0}
and also commute with all the ``local IM'' $I_{2k-1}$

\eqn\ti{[I_{2k-1},Q_n]=0\ . }
The operators $Q_n$ invariantly act on each of the level subspaces
${\cal F}_{p}^{(l)}$; in particular, the vacuum state $\mid p \rangle$
is the eigenstate of all $Q_n$,

\eqn\li{Q_n\mid p \rangle = Q_{n}^{(vac)}(p)\mid p \rangle\ ,}
where the eigenvalues $Q_{n}^{(vac)}(p)$ are given by the integrals

\eqn\mcnaa{\eqalign{
Q_{n}^{(vac)}(p)&=\int_{0}^{2\pi}du_1 \int_{0}^{u_1}dv_1
\int_{0}^{v_1}du_2 \int_{0}^{u_2}dv_2 ...
\int_{0}^{v_{n-1}}du_{n}\int_{0}^{u_n}dv_n\cr
\prod_{j>i}^{n}\Biggr[&{\Bigl(2\ \sin\big({{u_i - u_j}\over 2}\big)
\Bigl)}^{2\beta^2}\
{\Bigl(2\ \sin\big({{v_i - v_j}\over 2}\big)\Bigl)}^{2\beta^2}\Biggr]
\prod_{j\geq i}^{n}{\Bigl(2\ \sin\big({{u_i - v_j}\over 2}\big)
\Bigl)}^{-2\beta^2}\cr
\prod_{j>i}^{n}&{\Bigl(2\ \sin\big(
{{v_i - u_j}\over 2}\big)\Bigl)}^{-2\beta^2}
\ \ 2\ \cos\Bigl(2 p\ \big(\pi+\sum_{i=1}^{n}
(v_i - u_i ) \big)\Bigl)\ .}}
In particular

\eqn\tur{Q_{1}^{(vac)}(p) = {{{4{\pi}^2}\ \Gamma (1-2\beta^2)}\over
{\Gamma(1-\beta^2-2 p)\ \Gamma(1-\beta^2+2    p)}}\ .}

Using the power series expansion \ \bcvf\  one can show that the operator
${\bf T}(\lambda)$ is entire function of $\lambda^2$ (just as it was in
the classical case) in the sense that all its matrix elements and its
eigenvalues $t(\lambda)$
are entire functions of this variable. In fact, the
eigenvalues $t (\lambda)$ exhibit essential singularity at the infinity
as the result of accumulation of zeroes of these functions along the
real axis in $\lambda^2$-plane as $\lambda^2\to -\infty$
\footnote{$^4 $}{Validity of this picture is restricted to the domain
\cent .}.
Asymptotic form of the
operator ${\bf T}(\lambda)$ at $\lambda^2 \to \infty$ is of primary
interest because,
in view of \ \qexp ,  it is in this limit one anticipates
to get in touch with
the ``local IM''\  \inas, \  \pol . Rough estimate of the
matrix elements of the
operators \ \plo\  along
the lines proposed in \ \FSLS\  as well as physical
arguments based on the thermodynamic treatment of the ``Coulomb gas
partition function'' \ \mcnaa\  gives the leading asymptotic

\eqn\zor{\log[{\bf T}(\lambda)]\sim  m\ \lambda^{1+\xi}\ ,}
where
\eqn\ju{\xi={\beta^2\over 1-\beta^2}\ .}
The constant $m$ is expected to depend on $c$ but not on the
particular matrix element of this operator; this asymptotic form
holds in the domain
\eqn\cnn{{-\pi+\epsilon < {\rm arg}\big(\   \lambda^2\
\big) < \pi - \epsilon},
\qquad {\lambda \to \infty}}
with arbitrary small positive $\epsilon$. So far we could not find any
direct way to compute the constant $m$. Therefore we used the Algebraic
Bethe Ansatz\ \FadS\  to study
the discrete approximations to the integral in
\ \loi ; in taking the continuous limit the integral form of the Bethe
Ansatz equations proposed
in \ \dVega\  is particularly useful. This way we
obtain

\eqn\der{{\bf T}(\lambda) =
\Lambda(q\lambda)+\Lambda^{-1}(q^{-1}\lambda)\ ,}
where $\Lambda(\lambda)$ expands in the domain \ \cnn\  into
the asymptotic series

\eqn\gl{
\log\Lambda(q\lambda)\simeq  m\ \lambda^{1+\xi}-\sum_{n=1}^{\infty}\ C_n\
I_{2n-1}\ \lambda^{(1-2n)(1+\xi)}\ ,}
where

\eqn\coef{\eqalign{&m={2 \sqrt \pi \Gamma\big({1\over 2}-{\xi\over 2}\big)
\over \Gamma\big(1-{\xi\over 2}\big)}
\ \Biggl(\Gamma\big({1\over 1+ \xi}\big)\Biggl)^{1+\xi}\ ,\cr
C_n={1+\xi\over  n!}&\ \Biggl({\pi\xi\over 1+ \xi}\Biggl)^n\
\Biggl(  {2\ \Gamma\big({1\over 2}-{\xi\over 2}\big)
\over m\  \Gamma\big(1-{\xi\over 2}\big)}
\Biggl)
^{2n-1}{\Gamma\big((n-{1\over 2})
(1+\xi)\big)\over\Gamma\big(1+(n-{1\over 2})\xi\big)}\  ,}}
and $I_{2k-1}$ are exactly the ``local IM'' \ \inas ,\ \pol .
The coefficients \ $C_n$\ here are closely
related to the canonical on-shell normalization of the local IM of
the quantum Sine-Gordon theory\ \Zamt .
Note that
the coefficients $c_n$ in \qexp\  can be recovered in the classical limit
$\xi \to 0$ ($c \to -\infty$).
Appearance of the
fractional powers of $\lambda$ in \ \gl\  should not be very surprising -
the exponentials in \ \loi\  have anomalous dimensions and so the spectral
parameter $\lambda$ is to be thought of as carrying the dimension
$[\ length\ ]^{-{1\over {1+\xi}}}$. The expansions \ \bcvf\
and \ \gl\  give highly
nontrivial analytic relation between the ``nonlocal IM'' \ \plo\
and ``local IM'' \inas ,\ \pol .

The higher spin operators ${\bf T}_j (\lambda)$ also admit the power
series expansion similar to \ \bcvf , the coefficients being of course
algebraically dependent from the operators\ $Q_n$\
in \ \bcvf . For the first few
coefficients we find

\eqn\vbm{\eqalign{
{\bf T}_j (\lambda) =& {\sin\big(2 \pi   P\  (2j+1)\big)
\over \sin\big(2 \pi
P\big)}+\lambda^2\  A_j\big(2\pi  P,\pi\beta^2\big)\  Q_1+\cr
&\lambda^4\  \Bigr[\  A_j(2\pi  P,2\pi\beta^2\big)\  Q_2+
B_j\big(2\pi   P,\pi \beta^2\big)
\ Q_1^2\ \Bigr]+
O(\lambda^6)\  ,}}
where numerical coefficients read explicitly
$$A_j(x,a)={1\over 4\sin x \sin a}\Biggr({\sin (2j+1)(x-a)\over\sin (x-a)}-
{\sin (2j+1)(x+a)\over\sin (x+a)}\Biggr)\ ,$$
$$\eqalign{
B_j(x,a)={1\over 16 \sin x \sin a \sin 2 a}\Biggr(&{\sin (2j+1)(x-2a)\over
\sin (x-a)\sin(x-2a)}+\cr
&{\sin (2j+1)(x+2a)\over
\sin (x+a)\sin(x+2a)}-
{2\ \sin (2j+1)x\  \cos a\over \sin (x-a)\sin (x+a)}\Biggr)\ .}$$
The operators $Q_1$ and $Q_2$ in
are the same as in \ \plo . This algebraic dependence
can be summarized by the functional relations

\eqn\dfrs{{\bf T}_j
 (q^{{1\over 2}}\lambda){\bf T}_j (q^{-{1\over 2}}\lambda) =
1+{\bf T}_{j-{1\over
2}}(\lambda) {\bf T}_{j+{1\over 2}}(\lambda)\ .}
Note that these relations are identical
to the functional relation
obeyed by the commuting transfer-matrices of the
integrable $XXZ$
model \ \KR ,\ \BazR ,\ \KluP\ . This is well expected as our
\ ${\bf T}$- operators appear to be the continuous field
theory versions of the lattice transfer-matrices.
The relations\ \dfrs\ can be derived from\ \yab\
essentially the same way they are obtained in the lattice theory, by using
the R-matrix fusion procedure\ \Rks  .
The simple form of \ \dfrs\ is due to the fact that
quantum determinant \ \Kor\ of the operator\
${\bf L}_{\frac{1}{2}}(\lambda)$
\ \loi\   is equal to 1.
Using the relation\ \dfrs\ one can show that
all the operators\  ${\bf T}_j (\lambda)$\  with\  $j\geq 1$\  are
also entire
functions of $\lambda^2$ as well as \  ${\bf T}(\lambda)$.

At generic values of $c$ the relations \ \dfrs\  just allow one to express
the higher ${\bf T}$ operators in terms of the lower ones. However it is
well known in the lattice theory that at particular values of the
parameters, when the $XXZ$ system can be reduced to
the $RSOS$ model\ \Abf ,
the functional relations truncate to become a finite system of
functional equations\ \BP ,\
\BazR . Of course, similar phenomenon happens in our
continuous theory. Restricting our attention to the domain \ \cent\
we find
that the most simple  truncation occurs at

\eqn\cds{c=1-3{{(2n+1)^2}\over {2n+3}},\ \ \ \xi={2\over 2n+1}
;\qquad n=1, 2, 3, ...\  .}
For given $n$ in \ \cds\  consider
the finite collection of the Fock spaces

\eqn\kiop{{\cal F}_{p_k}\  ;\ \  p_k={2k-2n-3\over   2\   (2 n+3)}\ ,
\qquad k= 1, ..., n+1\ .}
At these values of $p$ the Fock spaces \ \kiop\  are known to be reducible
with respect to the action of the Virasoro algebra (they correspond to
the $(1,k)$ degenerate representations in the Kac classification
\ \Kac ,\  \Fei ).
Let us denote ${\cal V}_{p_k}$ the associated irreducible Virasoro
module obtained from \ \kiop\  by factoring out all
the submodules. Then the
space

\eqn\mol{
{\cal H}_{chiral} ({\cal M}_{2, 2n+3})=\oplus_{k=1}^{n+1} {\cal V}_{p_k}}
coincides with the space of chiral states of the ``minimal CFT'' ${\cal
M}_{2, 2n+3}$. It is possible to show that the operators ${\bf T}_j
(\lambda)$ with $j=0, {1\over 2}, 1, ..., n+{1\over 2}$ invariantly act
in the space \ \mol\  and being restricted to this space these operators
satisfy the symmetry relation

\eqn\juy{{\bf T}_{n+{1\over 2}-j}(\lambda)={\bf T}_j (\lambda);\quad j=0,
{1\over 2}, ..., n+{1\over 2}\ ,}
in particular ${\bf T}_{n+{1\over 2}}(\lambda)={\bf I}$. Under these
circumstances the relations \ \dfrs\  become a finite system of functional
equations

\eqn\kij{
\eqalign{ &t_j (q^{{1\over 2}}\lambda)\
t_j (q^{-{1\over 2}}\lambda) = 1+ t_{j+{1\over
2}}(\lambda)\  t_{j-{1\over 2}}(\lambda)\ ;\cr
& t_0 (\lambda) =  t_{n+{1\over 2}}(\lambda)=1\ ;
\qquad  t_{n+{1\over 2}-j}(\lambda)= t_j (\lambda)\ ,}}
which is satisfied by all eigenvalues $t_j (\lambda)$ of the operators
${\bf T}_j (\lambda)$
in the space\  \mol . We conjecture that any solution
$t_j (\lambda)$ to the functional equations \ \kij\  which is an entire
function of $\lambda^2$ and has the asymptotic behavior \ \zor\  in the
domain\  \cnn\  corresponds to some
eigenstate of ${\bf T}_j (\lambda)$ in the space \ \mol .

It was recognized before\ \KluP\  that the substitution

\eqn\bcv{Y_j (\theta) = t_{j+{1\over 2}}(\lambda)\
t_{j-{1\over 2}}(\lambda)\ ;
\qquad \lambda=\exp\bigl({\theta\over 1+\xi}\bigl)}
brings the system \ \kij\  to the form

\eqn\mwer{\eqalign{&
Y_j \big(\theta + {{i\pi\xi}\over 2}\big)\
Y_j \big(\theta - {{i\pi\xi}\over 2}\big)=
\big(1+Y_{j+{1\over 2}}(\theta)\big)
\big(1+Y_{j-{1\over 2}}(\theta)\big)\ ;\cr
&Y_0 (\theta)=Y_{n+{1\over 2}}(\theta)=0\ ;\qquad Y_{n+{1\over
2}-j}(\theta)=Y_j (\theta)\ ,\ }}
which coincides with the functional form of the Thermodynamic Bethe
Ansatz (TBA) equations\ \Zams\
for the ``massless S-matrix'' theory associated
with the minimal
CFT ${\cal M}_{2,2k+1}$\ \Klassen . In this ``massless TBA''
approach one assumes that the states of the CFT ${\cal M}_{2,2k+1}$ {\it
in infinite volume} can be interpreted as the scattering states of a
collection of $n$ massless right-moving particles
$A_j;\quad j={1\over 2}, 1, {3\over 2}, ... , {n\over 2},$ with the
energy-momentum spectrum $E=P$. Parameterizing the energy-momentum for
the particle $A_j$ in terms of rapidity $\theta$ as

\eqn\energ{E_j (\theta)=P_j (\theta)= \frac{m_j}{2}\ e^{\theta}\ ,\ \ \ \
m_j=\frac{2 m}{\pi} \  \cot\Bigr({\pi \xi\over 2}\Bigr)\
\sin\big(\pi\xi \ j\big)\ .}
one conjectures  the purely elastic and factorizable
S-matrix for these particles with the two-particle elements
$S_{j  j'}(\theta - \theta')$ (describing the scattering
processes $A_{j}
(\theta)+A_{j'} (\theta') \to A_{j} (\theta)+A_{j'} (\theta')$,
in obvious
notations) given by\  \FRK ,\ \Sm
\eqn\smatr{S_{j  j'}(\theta)=F_{j+j'}(\theta)
F_{|j-j'|}(\theta)
\prod_{k=1}^{2 \min (j,j')-1} F^2_{|j-j'|+k}(\theta)\ ,}
where the notation
$$ F_x(\theta)={\sinh \theta+i \sin(\pi\xi\  x)\over
\sinh \theta-i \sin(\pi\xi\  x)}\ $$
is used.
The S-matrix \ \smatr\  allows one to determine
the spectral density of states
in this infinite-volume system and hence to compute the free energy of
this system at finite temperature\  $ R^{-1}$. As the
thermal ensemble corresponds to the circular compactification of
``imaginary time'' one can interchange the roles of the space and the
``imaginary time'' and reinterpret this free energy as the ground-state
energy $e_0 (R)$ of the finite-volume
system with periodic boundary condition
defined on the circle with the circumference $ R$ \footnote{$^5$}{Of
course $R$ can be arbitrarily changed by the scale
transformation which is the symmetry of CFT. Note that in our previous
discussion of the ${\bf T}$-operators
this parameter was set equal to $2\pi$.}. This way one
obtains

\eqn\hdgf{e_0 (R)=-
\sum_{j}\ \frac{m_j}{2}\ \int_{-\infty}^{\infty}
{{d\theta}\over
{2\pi}}\
e^{\theta}\   \log(1+e^{-\epsilon_{j}(\theta)})\ ,}
where the functions $\epsilon_j (\theta)$ solve the integral equations
of TBA:
\eqn\inteq{
\frac{R m_j }{2}\
e^{\theta}
=\epsilon_j (\theta) + \sum_{j'}\int_{-\infty}^{\infty}
{{d\theta'}\over
{2\pi}}\ \varphi_{j
j'}(\theta -\theta')\  \log(1+e^{-\epsilon_{j'}(\theta')})
\ .}
with the kernel

$$\varphi_{j j'}(\theta)=-i\  \partial_{\theta} \ln S_{j j'}(\theta)\ .$$
By inverting
the integral operator in the right hand side of \ \inteq\  one
can show, following \ \Zams , that the functions

\eqn\liko{Y_{j}^{gr.st.}(\theta)= e^{\epsilon_{j} (\theta)}}
satisfy the functional equations \ \mwer\  and agree
with the asymptotic condition
\ \zor\ provided the parameter \ $m$\ in\ \energ   \ is the
same as in\ \zor . The ground state (the state of
the lowest eigenvalue of $L_0$) of the CFT
${\cal M}_{2,2n+3}$ is the vacuum state $\mid p_{n+1}\rangle$ from \ \mol .
We have checked (for the case of ${\cal M}_{2,5}$) that for
this value of $p$ the
numerical solution of the integral equations \ \inteq\
(which is named the ``kink solution'' in \ \Zar ) matches perfectly both
the expansions \ \gl\  and \ \vbm \ \BASZ .
We see that the functions $\epsilon_j
(\theta)$ of TBA admit an interpretation in terms of the eigenvalues of
the ${\bf T}$ -operators in the ``cross-channel''.

Let us stress that while the solution to
the integral equation \ \inteq\  is
unique,
the functional equations \ \mwer\  admit infinitely many solutions
even in the class of entire functions with the asymptotic behavior
\ \zor . From the analytic point of view these solutions (which are in
one-to-one correspondence with the eigenstates of the ${\bf T}$
operators in the space \ \mol , by our conjecture) can be characterized by
the patterns of zeroes of the functions $t_j(\lambda)$
in the complex $\lambda^2$
plane. All the solutions have infinitely many zeroes at the negative
segment of the real axis in this plane (with $-\infty$ as their
accumulation point) and some finite number of zeroes away from this
locus. It is characteristic feature of the ground state solution \ \liko\
that $all$ zeroes of the functions $t_{j}^{gr.st}(\lambda)$ are located
at $\Im m\lambda^2=0;\  \Re e
\lambda^2 < 0$; under this condition \ \inteq\  follow from
\ \mwer . The eigenvalue functions
$t_j (\lambda)$ associated with the other
eigenstates in \ \mol\  have more complex patterns of zeroes. For example,
for the vacuum states
$\mid p_k \rangle ;\  k= 1, 2, ..., n$ the eigenvalue
functions $t_j (\lambda)$ have finite number of
zeroes at $\Im m\lambda^2=0;\
\Re e\lambda^2 > 0$
and no complex zeroes, and for the excited states in \ \mol\
these functions exhibit also complex zeroes which can be interpreted as
the rapidities of the massless particles in these states
\footnote{$^6$}{It would be extremely interesting to find out how the
classification of states by the patterns of zeroes relates to the
``fermionic sum representations'' of the Virasoro characters
\ \MKlas  .}. In all cases
one can use the technique developed in \ \KluP\  to convert the functional
equations \ \mwer\  into integral equations similar to \ \inteq\  but with
additional terms in the left-hand sides. We have analyzed numerically
these ``excited states TBA equations'' for few simplest states in the
CFT ${\cal M}_{2,5}$ and found again a perfect agreement with the
expansions \ \gl\  and\  \vbm \ \BASZ .

Strictly speaking, the above discussion concerned the case of
CFT and associated massless TBA. It is known however that the
integrable structure of CFT remains essentially
intact in more general massive
Quantum Field Theory obtained by perturbing this CFT with the relevant
primary field $\Phi_{1,3}$\ \ABZ . Namely, this perturbed field theory
exhibits two infinite sets of commuting local Integrals of Motion,
$I_{2k-1}$ and ${\bar I}_{2k-1}$, $k= 1,2, ...$, where $I_{2k-1}$ are
obtained by appropriate deformations of the ``right'' conformal IM\ \inas\
and ${\bar I}_{2k-1}$ are the deformations of the same IM from the
``left'' sector of the CFT (see \ \ABZ ). There are good reasons to believe
that the above operators ${\bf T}_j (\lambda)$, with suitable
deformations\footnote{$^7$}{In fact, it is not very difficult to figure
out what these ${\bf T}$-operators in the perturbed theory should look
like, if one recalls the known relation between the integrable
structures of KdV and sine-Gordon theories\ \Fa ,\  \jap ,\
\Eguchi ,\ \Kup\   and the relation between the
quantum Sine-Gordon theory and the perturbed minimal CFT explained in
\ \Leclair, \Sm .
As we did not yet elaborate all the details we do not present these
deformed ${\bf T}$-operators here.}, can be extended to the case of the
perturbed theory in such a
way that they satisfy the equations \ \mcn  \ and \ \dfrs .
They also enjoy the
same asymptotic expansion \ \gl\  in terms of (deformed) ``right'' IM
$I_{2k-1}$. The major difference is that the deformed operators ${\bf
T}_j (\lambda)$ are no longer entire functions of $\lambda$; instead
they are expected to have an essential singularity at $\lambda = 0$
(being regular everywhere else in the finite part of the $\lambda$
plane) which is controlled by the asymptotic expansion in terms of the
``left'' IM ${\bar I}_{2k-1}$, similar to \ \gl , with $\lambda$ replaced
by $\mu / \lambda$, where $\mu$ is related in a simple way
to the coupling constant
of the perturbed theory.
In the case of perturbed minimal CFT
${\cal M}_{2,2n+3}$ the functional relations \ \dfrs\  become again the
closed set of functional equations \ \kij ,
but now one has to look for the
solutions which have essential singularity at $\lambda = 0$ and enjoy
the asymptotic behavior

\eqn\xz{ \log {\bf T}_j (\lambda)
\sim \cases{m \ \lambda^{1+\xi},\ & if \
$\lambda \to \infty$\ ; \cr m\ \big({\mu \over
\lambda}\big)^{1+\xi},\ & if  \ $\lambda \to 0$\ .\cr}}
Like in the case of CFT, it is very plausible that there is one to one
correspondence between the solutions of \ \kij\  with these analytic
characteristics and the stationary states of the perturbed CFT ${\cal
M}_{2,2n+3}$ on finite circle. We studied
numerically \ \BASZ\  few simplest
solutions in the case of perturbed ${\cal M}_{2,5}$, again with the
excellent agreement with the data available through the Truncated
Conformal Space method\ \Yur .

We can not resist the temptation to mention here a remarkable
relation between the operators \ \mjvb\  and CFT with non-conformal
interactions at the boundary. According to the analysis in
\ \Card , in a
minimal CFT (the one where the sum \ \jdgt\
contains finite number of terms)
with a boundary only finitely many {\it conformally invariant} boundary
conditions (CBC) is possible. These CBC are in one
to one correspondence with the terms in \ \jdgt ; we denote $B_a$ the
CBC associated with ${\cal V}_a$. The boundary state $\mid B_a
\rangle \in {\cal H}_{phys}$ describing CBC $B_a$ has the form

\eqn\bound{
\mid B_a \rangle = \sum_b {{{\bf S}_{a b}}\over {\sqrt{{\bf S}_{0 b}}}}
\sum_{\mu} \mid a_\mu \rangle \otimes \mid {\bar a}^{\mu}\rangle\ ,}
where ${\bf S}$ is the matrix of modular transformation of characters,
$\lbrace \mid a_\mu \rangle \rbrace$ is an arbitrary basis in ${\cal
V}_a$ and  $\lbrace \mid {\bar a}^\mu \rangle \rbrace$ is the dual basis
in ${\bar{\cal V}}_a$. Furthermore,
with a given conformal boundary condition $B_a$, the space of local
boundary operators\footnote{$^{8}$}{We call here {\it local} the boundary
operators which exist as the insertions at the points where two
components of the boundary with the {\it same}
boundary conditions meet, in
contrast with the operators associated with the juxtapositions of
{\it different} boundary conditions,
see \ \Card .} forms the subspace ${\cal
H}_{B_a}$ in ${\cal H}_{chiral}$ isomorphic to the direct sum
$\oplus_{b} {\cal V}_{b}$ where only those $b$ which have
nonvanishing fusion constants $N_{a a}^{b}$ are admitted. One can
obtain more general non-conformal boundary conditions by perturbing the
CBC $B_a$ with relevant operators $\psi_{b} \in {\cal H}_{B_a}$. Let us
denote $B_a (g)$ the CBC $B_a$ perturbed with the operator
$\psi_{1,3}$, the primary field associated with the highest weight
vector in ${\cal V}_{1,3}$; here $g$ stands for the
coupling parameter. The corresponding boundary state is

\eqn\der{\mid B_a (g)\rangle = {\cal P} \exp \big (
-g \int_{0}^{2\pi} d u\ \psi_{1,3}(u)  \big )\mid B_a \rangle\ ,}
where we have assumed the geometry of a half-infinite cylinder with $u$
being the coordinate along the boundary,
$\psi_{1,3}(u+2\pi)=\psi_{1,3}(u)$. It is known\ \GZ\  that minimal CFT
with the perturbed boundary condition $B_a (g)$ is
integrable. Now we recall\ \Card\
the standard correspondence between the vectors
in ${\cal H}_{chiral}\otimes {\bar {\cal H}}_{chiral}$ and endomorphisms
of ${\cal H}_{chiral}$ and denote ${\bf B}_j (g)$ the
operator in ${\cal H}_{chiral}$ associated with the boundary state \
\der\ with $a = (1, 2j+1)$.
Using \ \bound\ with the explicit form of the matrix ${\bf S}$
for the minimal model ${\cal M}_{n,n'}$ \
$(\ \beta^2=\frac{n}{n'},\  n'>n=2,3,...)$\  we obtain

\eqn\mcewq{{\bf B}_j (0)= \sum_{k,l}\ \Bigr({n n'\over 8}
\Bigr)^{-{1\over 4}}\
\Biggr[\frac{\sin\big(\frac{\pi}{2\beta^2}
\ p_{k,l}\big)} {\sin\big(2\pi\  p_{k,l}\big)}
\Biggr]^{\frac{1}{2}}
\ \sin\big(2\pi\ p_{k,l}\  (2j+1)
\big)\ {\bf P}_{k,l}\ ,}
where ${\bf P}_{k,l}$ are the projectors on the subspaces ${\cal
V}_{k,l}$ and

$$2\ p_{k,l}=\beta^2\ l-k$$
are the eigenvalues of the operator $P$ on these subspaces. Comparing
this expression with \ \vbm\  we see that

\eqn\uy{{\bf B}_j (0) = \Bigr({{n n'}\over 8}\Bigr)^{-{1\over 4}}\
\Bigr(\sin\big(2\pi P\big)\
\sin\big(2\pi\beta^{-2} P\big)\Bigr)^{\frac{1}{2}}\ \ {\bf T}_j (0)\ .}
Of course this relation is not very surprising as the operators ${\bf
T}_j (0)$ are essentially the Verlinde operators $\phi_a$\
\Ver\  for $a=(1,
2j+1)$ and the eigenvalues of the latter are known to be
${\bf S}_{a b} / {\bf S}_{0 b}$. What is less trivial is that the
$\cal P$-exponential in \ \troi\  gives
exactly the Feigin-Fuchs realization for the  $\cal P$-exponential in
\ \der\  and so this relation extends to the case of full boundary state
${\bf B}_j (g)$, i.e.

\eqn\uyas{{\bf B}_j (g) = \Bigr({{n n'}\over 8}\Bigr)^{-{1\over 4}}\
\Bigr(\sin\big(2\pi P\big)\
\sin\big(2\pi\beta^{-2 }P\big)\Bigr)^{\frac{1}{2}}\ \ {\bf T}_j (\lambda
)\ ,}
where the relation between $\lambda$ and $g$ depends
on the normalization of the boundary field $\psi_{1,3} (u)$ in
\ \der . Using the operator product expansion
$$\psi_{1,3}(u) \psi_{1,3}(u') = (u-u')^{-2 \Delta_{1,3}} +
\ \ {\rm  less \ singular \ terms} $$
to fix this normalization we find
\eqn\mvfep{g^2 = \lambda^4\  {{\sin\big(2\pi(j+1)\beta^2 \big)
 \sin\big(2\pi j \beta^2
\big)}\over {\pi \sin \big(2\pi\beta^2 \big)}}\ {{\Gamma^3 (1-\beta^2 )
\Gamma(3\beta^2 -1)}\over {1-2\beta^2}}.}
The relation \ \uyas\  allows
one to interpret many properties of the ${\bf
T}$-operators discussed in this paper in terms of Renormalization Group
flows between different CBC. We believe it also throws new light on the
functional equations for the boundary
partition functions obtained in\ \FSLS\
by TBA approach. More details about this relation will be given in \ \BASZ.

The above discussion was restricted to the domain \ \cent\
of $c$ as for $c
> -2$ the integrals in \troi\  become divergent. However this limitation is
not very significant. It is possible to change the integration contours
in \ \troi\  in such a way that the operator
\ \loi\  will make sense for any $c$
(this can be thought of as analytic continuation in $c$). Moreover, with
this redefinition most of the above properties of the ${\bf T}_j(\lambda)$
operators will still hold in larger domain
$-\infty < c < 1$, the most significant change being in the nature of
the essential singularity of ${\bf T}_j (\lambda)$ at $\lambda =
\infty$ for $-2 < c < 1$ ( \ \gl\   remains valid but in the
smaller domain $-2\pi (1 - \beta^2 ) < {\rm arg}\big(\
\lambda^2\ \big) < 2\pi (1 - \beta^2)$ while completely different
asymptotic behavior emerges in the complimentary sector).
 Although we did not yet completed the analysis
for this domain $-2 < c < 1$, for the cases of unitary minimal models
${\cal M}_{n, n+1}$ \ \Zatr\  we expect to observe the ``truncation'' of
the functional relations \ \dfrs\   similar to the one for
${\cal M}_{2, 2n+3}$\ discussed above.
We hope to return to this point elsewhere.

And, of course, the above approach can be generalized
in a straightforward way to describe the quantum theory
of generalized KdV\ \Drin\  associated with
the higher-rank simply laced Lie algebras, or
equivalently, to describe the integrable structure of
the CFT with the extended W-symmetry.

\centerline{}
\centerline{}
\centerline{{\bf Acknowledgments}}

\centerline{}

V.B. thanks R.J. Baxter for interesting discussions and
the Department of Physics and Astronomy, Rutgers University
for the hospitality.
S.L. and A.Z. are grateful to
V.A. Fateev and Al.B. Zamolodchikov for sharing their insights and
important comments. S.L. also acknowledges helpful discussions with
A. LeClair.

This work is supported in part by NSF grant (S.L.) and
by DOE grant $\#$DE-FG05-90ER40559 (A.Z.).
\listrefs

\end